\documentclass[sigconf]{acmart}

\usepackage{multirow}
\usepackage{color}
\usepackage{subcaption}
\usepackage{xspace}
\usepackage{enumitem}

\newcommand{\ie}{\emph{i.e.,}\xspace}
\newcommand{\eg}{\emph{e.g.,}\xspace}
\newcommand{\etal}{\emph{et al.}\xspace}

\AtBeginDocument{%
  }

\copyrightyear{2022} 
\acmYear{2022} 
\setcopyright{acmcopyright}
\acmConference[ICTIR '22]{Proceedings of the 2022 ACM SIGIR International Conference on the Theory of Information Retrieval}{July 11--12, 2022}{Madrid, Spain}
\acmBooktitle{Proceedings of the 2022 ACM SIGIR International Conference on the Theory of Information Retrieval (ICTIR '22), July 11--12, 2022, Madrid, Spain}
\acmPrice{15.00}
\acmDOI{10.1145/3539813.3545124}
\acmISBN{978-1-4503-9412-3/22/07}

\begin{document}

\title{Do Loyal Users Enjoy Better Recommendations? Understanding Recommender Accuracy from a Time Perspective}

\author{Yitong Ji}
\affiliation{%
  \institution{Nanyang Technological University}
  \country{Singapore}
}
\email{yitong.ji@ntu.edu.sg}

\author{Aixin Sun}
\affiliation{%
  \institution{Nanyang Technological University}
  \country{Singapore}
}
\email{axsun@ntu.edu.sg}

\author{Jie Zhang}
\affiliation{%
  \institution{Nanyang Technological University}
  \country{Singapore}
}
\email{zhangj@ntu.edu.sg}

\author{Chenliang Li}
\affiliation{%
  \institution{Wuhan University}
  \country{China}
}
\email{cllee@whu.edu.cn}

\begin{abstract}
In academic research, recommender systems are often evaluated on benchmark datasets, without much consideration about the \textit{global timeline}. Hence, we are unable to  answer questions like: \textit{Do loyal users enjoy better recommendations than non-loyal users?} Loyalty can be defined by the time period a user has been active in a recommender system, or by the number of historical interactions a user has.  In this paper, we offer a comprehensive analysis of recommendation results along global timeline. We conduct experiments with five widely used models, \ie BPR, NeuMF, LightGCN, SASRec and TiSASRec, on four benchmark datasets, \ie MovieLens-25M, Yelp, Amazon-music, and Amazon-electronic. Our experiment results give an answer ``No'' to the above question. Users with many historical interactions suffer from relatively poorer recommendations. Users who stay with the system for a shorter time period enjoy better recommendations. Both findings are counter-intuitive. Interestingly, users who have recently interacted with the system, with respect to the time point of the test instance, enjoy better recommendations. The finding on recency  applies to all users, regardless of users' loyalty. Our study offers a different perspective to understand recommender accuracy, and our findings could trigger a revisit of recommender model design. The code is available in \url{https://github.com/putatu/recommenderLoyalty}.
\end{abstract}



\maketitle

\section{Introduction}
\label{sec:introduction}

Recommender systems are often evaluated offline on a benchmark dataset in academic research~\cite{evaluatingRigorously, offlineOnline}. That is, a dataset is split into training set and test set. Then, a recommender model learns from the training set and is evaluated on the test set. Many papers report overall accuracy of their models on the test set, and conduct ablation studies. However, analysis of recommendation accuracy from time dimension is often lacking. In this paper, we study recommendation accuracy by time-related factors, and aim to answer questions like: \textit{Do loyal users enjoy better recommendations than non-loyal users?}

\begin{figure}
    \centering
    \includegraphics[width=0.95\columnwidth]{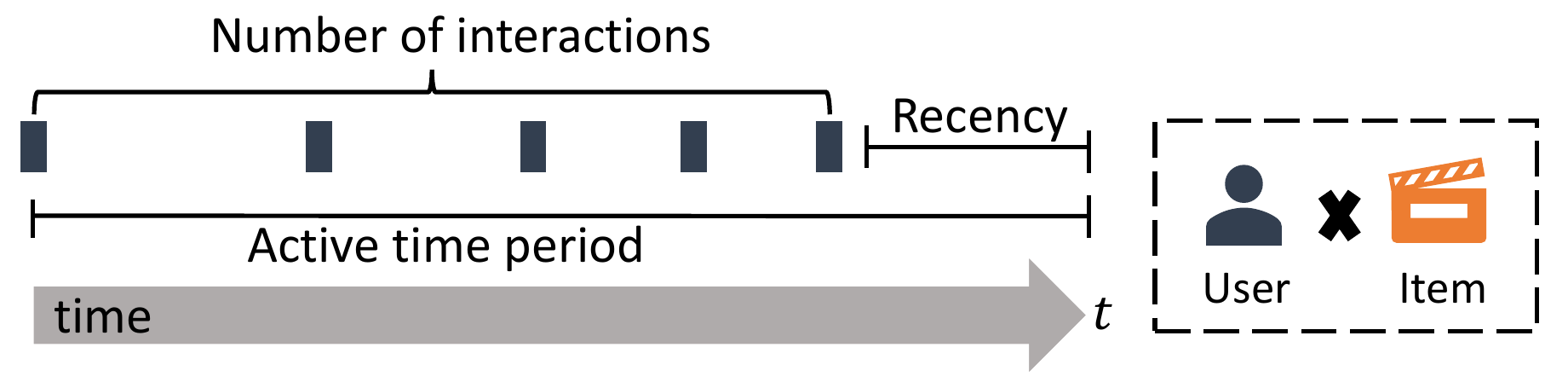}
    \caption{The time factors we considered in our analysis}
    \label{fig:timeFactor}
\end{figure}

We consider three time factors along the global timeline, illustrated in Figure~\ref{fig:timeFactor}. Suppose a test instance happens at time $t$, we have (i) \textbf{number of accumulated interactions} of a user before time point $t$; (ii) \textbf{active time period}, \eg number of days since the user's first interaction with any item in the system till time $t$; and (iii) \textbf{recency}, \eg number of days from the user's previous interaction to time $t$. We consider both ``number of accumulated interactions'' and ``active time period'' as loyalty indicators, because both factors reflect a user's relationship with the system, or to what extent the user is familiar with the system. 

We evaluate five recommendation models - BPR~\cite{BPR}, NeuMF~\cite{neuMF}, LightGCN~\cite{lightgcn}, SASRec~\cite{sasRec}, and TiSASRec~\cite{timeSasRec}. Among them, BPR, NeuMF, and LightGCN  are general batch mode recommenders without considering time in their design. That is, they model a user's preference using all his/her historical interactions, and do not consider at what time points these interactions take place. SASRec is a sequence-aware model which considers user's historical interactions in chronological order, while TiSASRec is a time-aware recommender which takes time into consideration in their model design. Our experiments are conducted on four publicly available datasets, all with 10 years of interactions: MovieLens-25M, Yelp, Amazon-music, and Amazon-electronic.

Experiment results show that \textit{loyal users suffer from poorer recommendations}, compared to non-loyal users for general recommenders BPR, NeuMF, and LightGCN. This finding is counter-intuitive. In general, we expect that preferences of loyal users can be more accurately modeled by recommenders, because they have more interactions or have spent more time with the system. This counter-intuitive finding provokes us into thinking whether \textit{all historical interactions of users} are indeed useful for learning user preference. For sequential recommenders, \ie SASRec and TiSASRec, there are cases whereby loyal users and non-loyal users enjoy comparable recommendation accuracy. One reason is that SASRec and TiSASRec work with each user's local timeline, hence they treat recent interactions of a user differently from his/her past interactions. Here, local timeline refers to the sequential order of a specific user's interactions, without considering the absolute time of these interactions along the global timeline.

Further analysis on the \textbf{recency} factor shows that, users who have interactions close to the test time point $t$, receive better recommendations, than those who last interact with the system some time ago. This observation holds for all the five recommenders, regardless of whether they are time-aware or not. It is a strong evidence that more recent interactions, with respect to the test time, are more helpful in recommendation. Here, recent interactions are defined with respect to the \textit{global timeline} (see Figure~\ref{fig:timeFactor}). Note that, time-aware models SASRec and TiSASRec do not show better ability in distinguishing recent interactions along the global timeline, because they model a user's interactions as a sequence and does not consider interactions with respect to the test time.

Our research findings suggest that user's interests change in a long run. More recent interactions better reflect users' current interests, while past interactions could be considered outdated. However, recent interactions refer to a dynamic and changing set of interactions along the global time dimension because the number of available items in the system may change from time to time. These interactions cannot be accurately modeled based on a user's local timeline. That is, to a specific user, the most recent interaction with respect to his/her past interactions may not be ``recent'' with respect to a time point along the global timeline. For example, a user's last purchase from an e-commerce website may happen in last day, last week, or even last year. Based on our findings, we call for a revisit on recommender model design. Rather than treating all historical interactions equally, we might need to pay more attention to recent interactions when a recommendation decision is to be made, along the global timeline. 

\section{Related Work}
\label{sec:related_work}

Many existing recommender models take time into consideration. Commonly, time is taken into account to preserve sequential order of a user's interactions. For instance, GRU4REC~\cite{GRU4REC} applies Recurrent Neural Network (RNN) on sessions to capture sequential dependencies between items in user's interactions. Some other studies~\cite{NARM, STAMP} combine attention mechanism with RNN network, and model not only the sequential relationship in sessions but also relative importance of items along the sessions. Inspired by~\textit{Transformer}, self-attention network based recommenders~\cite{sasRec, timeSasRec, seqHypergraph, intentRec} have also been widely adopted to learn long-term dependencies. Positional embedding is used in self-attention network to model position of an item in an interaction session. Time has also been considered in recommendation as an attribute or contextual information. Analysis on Amazon-datasets shows that demand of T-shirts follows a seasonal cycle~\cite{timeMatters}. Hansen \etal~\cite{contextMusicRec} highlight the importance of time context in music consumption. There is also study on the time context in emoji recommendation~\cite{Caper}. These recommenders embed time context into embedding vectors and incorporate time context features in recommendation. Another way to take time into consideration is by relative time. A few studies~\cite{timeMatters, timeSasRec} emphasize the importance of time interval between interactions in recommendation. 

Time has also been factored in through recency of past interactions, particularly in the news recommendation task. Recency is defined with respect to the time when a recommendation is made. The authors in~\cite{recencyReleancy} acknowledge the importance of recency, and propose a news recommender model that balances the trade-off between recency and relevancy. Here, recency concerns \textit{item}'s age at the time point of recommendation, as the items are news articles. Hence, their definition of recency is different from ours. 
Other work like~\cite{goodTimeBadTime} considers recency of past interactions in collaborative filtering. It shows that adopting different recency windows in recommender design can lead to different recommendation accuracy. Similarly, the authors of~\cite{recencyCF} demonstrate the influence of recency based decay functions used in collaborative filtering recommenders. 

The above reviewed papers incorporate time factors in their recommender designs, either motivated by pre-experiment analysis on datasets, or by intuition and domain knowledge. In contrast, we conduct \textit{post-experiment analysis} on the recommendation results of various models, on multiple datasets representing different recommendation tasks. Our analysis further explains the influence of time factors (\eg active time period, and recency) in general recommendation setting.

\section{Experiment}
In this section, we detail datasets, the evaluated recommendation models, and the experiment setup.

\subsection{Datasets and Evaluated Models}
\label{ssec:dataset}

\begin{table*}[t]
  \begin{center}
  \small
    \caption{Statistics of the four datasets, all covering interactions in 10-year period from their starting time.}
    \label{tab:datasetStats}
    \begin{tabular}{l|ccrrr} 
      \toprule
      Dataset & Starting time & Data filtering & \#User & \#Item& \#Data instances \\      
      \midrule
      MovieLens-25M & 21 Nov 2009  & No filtering & $62,202$ & $56,774$ & $9,808,925$ \\
      Yelp & 13 Dec 2009 & 10-core & $116,655$ & $61,027$ & $3,127,215$ \\
      Amazon-music & 02 Oct 2008 & 5-core & $11,651$ & $9,243$ & $114,833$ \\
      Amazon-electronic & 05 Oct 2008 & 10-core & $109,990$ & $39,552$ & $1,752,238$ \\
      \bottomrule
    \end{tabular}
  \end{center}
\end{table*}

\begin{table*}[t]
  \begin{center}
  \small
    \caption{Average \#Interactions, ATP (days) and Recency (days) for each user group.}
    \label{tab:distribution}
    \begin{tabular}{l|l|cc|cc|cc} 
      \toprule
      \multirow{2}{*}{Dataset}& \multirow{2}{*}{Baselines} & \multicolumn{2}{c|}{\#Interactions} & \multicolumn{2}{c|}{ATP (days)}& \multicolumn{2}{c}{Recency (days)} \\
      & & Loyal Users & Non-loyal Users & Loyal Users & Non-loyal Users & Active Users & Inactive Users  \\
      \midrule
      \multirow{3}{*}{MovieLens-25M} & General Recommenders & $505.0$ & $61.4$ & $909.2$ &$3.3$ & $1.6e-4$& $32.2$\\
      & SASRec  & $40.0$ & $28.1$  & $385.9$& $1.1$& $1.6e-4$& $32.2$\\
      & TiSASRec  & $36$ & $26.4$  &$366.9$ & $1.0$& $1.6e-4$& $32.2$\\
      \midrule
      \multirow{3}{*}{Yelp} & General Recommenders & $48.2$ & $11.8$ & $2299.2$ & $964.1$ & $12.5$&$283.0$ \\
      & SASRec  & $17.9$ & $11.8$ & $1899.1$& $622.9$&$12.5$&$283.0$\\
      & TiSASRec  & $32.2$ & $11.8$ & $2173.8$& $864.9$&$12.5$&$283.0$ \\
      \midrule
      \multirow{3}{*}{Amazon-music} & General Recommenders & $17.7$ & $4.7$ & $1592.3$& $445.0$ & $1.7$&$447.8$ \\
      & SASRec  & $16.1$ & $4.7$  & $1578.5$& $434.2$& $1.7$&$447.8$\\
      & TiSASRec  & $15.6$  & $4.7$ & $1572.0$& $429.4$& $1.7$&$447.8$\\
      \midrule
      \multirow{3}{*}{Amazon-electronic} & General Recommenders & $21.8$& $10.2$ & $2208.5$& $1012.0$ & $15.1$ & $373.8$\\
      & SASRec  & $21.1$& $10.2$& $2188.9$& $1003.1$& $15.1$ & $373.8$\ \\
      & TiSASRec  & $18.5$ & $10.2$ & $2101.1$ &$940.1$ &$15.1$ & $373.8$\ \\
      \bottomrule
    \end{tabular}
  \end{center}
\end{table*}

We conduct experiments on four public datasets: MovieLens-25M, Yelp, Amazon-music, and Amazon-electronic. All interactions in these four datasets come with timestamps. From each dataset, we extract interactions in a 10-year time period for experiments (see Table~\ref{tab:datasetStats} for the starting time of each dataset). After that, we filter the extremely inactive items or users. We adopt $k$-core filtering for Yelp, Amazon-music, and Amazon-electronic. After $k$-core  filtering, all users and items in the dataset will have at least $k$ interactions. Depending on the dataset size and sparsity, we choose 10-core for Yelp and Amazon-electronic, and 5-core for Amazon-music. No filtering is performed on MovieLens-25M because MovieLens dataset ensures each user has at least 20 ratings.  Table~\ref{tab:datasetStats} reports the statistics of the four datasets after filtering.

Our findings are made on recommendation results by five widely used baseline models\footnote{We implement the general recommenders, \ie BPR, NeuMF and LightGCN, using Recbole~\cite{Recbole}. For SASRec and TiSASRec, we follow the implementations in \url{https://github.com/pmixer/SASRec.pytorch} and \url{https://github.com/JiachengLi1995/TiSASRec} respectively.}: BPR, NeuMF, LightGCN, SASRec, and TiSASRec. Each baseline represents one type of recommendation. BPR, NeuMF, and LightGCN are general recommender models which do not take time into consideration. \textbf{BPR} learns users' and items' latent factors via a pairwise ranking loss in matrix factorization. \textbf{NeuMF} learns user and item interaction function with a model that combines both matrix factorization and multi-layer perceptron. \textbf{LightGCN} is a graph-based recommender model that learns complex relationships between users and items, by using graph convolutional network. SASRec and TiSASRec are time-aware models which consider time at which interactions take place. \textbf{SASRec} is a self-attentive network that models the sequential pattern in user's behaviour. \textbf{TiSASRec} is inspired from SASRec but takes time intervals between events as input.

In our experiments, we adopt leave-one-out data split but with well consideration of data leakage, to be detailed in Section~\ref{ssec:expSetting}. For each model, we tune their hyperparameters by continuous random search, in each run of the experiment. Similar to other work~\cite{CTRec, knowledgeMemory}, we tune hyperparameters by using a validation set that consists of the second last interaction of each user. 

We adopt Hit Rate (HR) and Normalized Discounted Cumulative Gain (NDCG) as evaluation metrics. In this research, we rank \textbf{all available items} and make top-$N$ recommendations. That is, we do not use sampled metrics, because sampled metrics introduce bias in recommendation accuracy~\cite{sampledMetrics}.\footnote{In this paper, we only present the HRs and NDCGs on top-$10$ recommendations. Similar findings hold for top-$5$ and top-$20$ recommendations.}

\subsection{Evaluation Setting}
\label{ssec:expSetting}

Recently a few studies~\cite{dataLeakage, revisitOffline, temporalContext} highlight the issue of data leakage in offline evaluation of recommender system. In particular, data partition strategies (\eg leave-one-out and random-split-by-ratio) that do not follow global timeline will allow the training of recommendation model using \textit{future} training instances. Future training instances are the interactions that take place after the time point of a test instance. To strictly follow the global timeline and completely avoid data leakage, it is suggested to adopt a streaming setting~\cite{dataLeakage}, \ie taking interactions happened before a time point $t$ for training and predict test instances at $t$, with a changing $t$ along the global timeline. This streaming setting requires the recommenders to be designed to take in incremental inputs along timeline. However, BPR, NeuMF, LightGCN, SASRec and TiSASRec are not incremental. We understand incremental models exist. However, batch recommendation models remain the mainstream in academic research. Here, our objective is to conduct post-experiment analysis of batch models' results by time factors.

To minimize the impact of data leakage on batch models, we follow the evaluation setting used in~\cite{dataLeakage} for our experiments. The key idea is to keep the number of future instances reasonably small. Specifically, given a dataset, we first conduct \textit{leave-one-out} split. That is, for each user, we treat his/her last interaction as test instance, and the remaining interactions as training instances. We adopt \textit{leave-one-out} split in this study because it is a widely used data partitioning strategy in recommender system~\cite{evaluatingRigorously}. Moreover, it allows the recommendation models to have a complete picture of a user's historical interactions by masking only the last interaction of the user as test instance. After obtaining training and test sets, we follow~\cite{dataLeakage} to conduct evaluation on test instances that happen in a specific year. 

Recall that each dataset contains 10-year instances, we select test instances that happened in a particular year for evaluation. Assuming we take year-6, denoted by $Y6$, as the test year, then test instances that happened in $Y6$ will be in the test set for this run of experiment. All instances before $Y6$ (\ie $Y1$ to $Y5$) and the training instances in $Y6$ will be used for training. Based on this experiment setting, we run three sets of experiments on each of the four datasets, using $Y6$, $Y8$, and $Y10$ as test years respectively. We experiment on three test years because recommender behaviours can vary along the timeline~\cite{timeEval}. Through experiments, we find that results on three test years show very similar trends. Hence, we only present the results for test year $Y10$ to avoid having too many similar plots.

\section{Experiment Results}
\label{sec:expResults}

We now study recommendation results to answer questions on user loyalty. Here, loyalty can be indicated by: (i) number of accumulated interactions a user has, and (ii) active time period of a user. 

\subsection{Loyalty by Number of Interactions}
\label{ssec:numInteractions}

\begin{figure}
    \centering
\begin{subfigure}{\columnwidth}
        \centering
        \includegraphics[width=0.7\columnwidth]{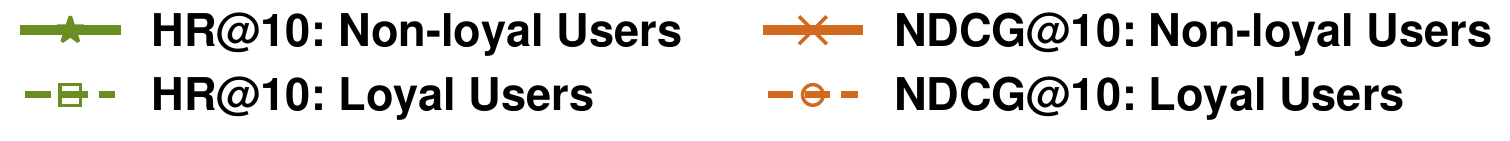}
        \Description{}
    \end{subfigure}
    \quad
    \begin{subfigure}{\columnwidth}
        \centering
        \includegraphics[width=\columnwidth]{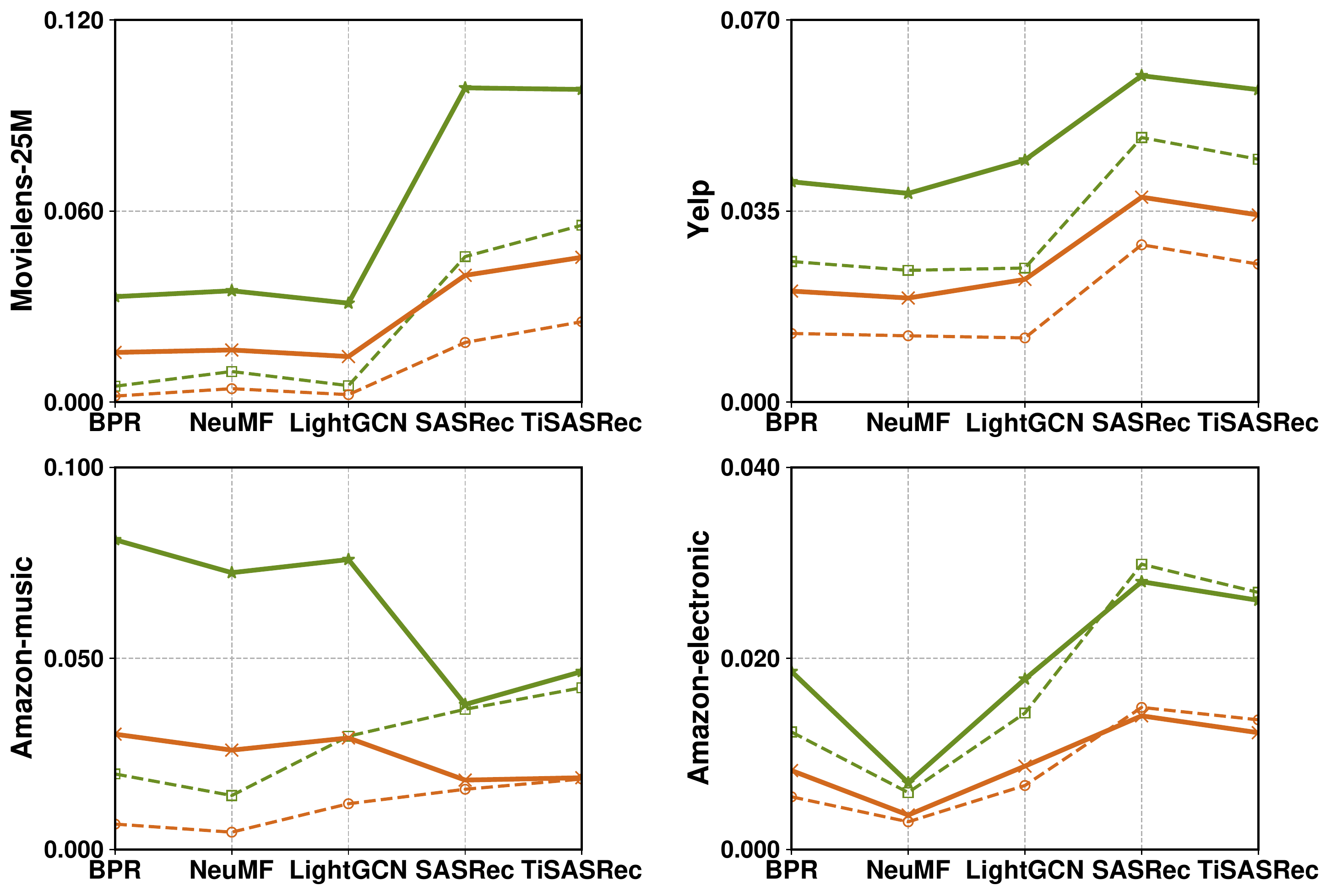}
    \end{subfigure}
    \caption{HR@10 and NDCG@10 of loyal and non-loyal users by \textbf{number of interactions}, in test year $Y10$}
    \label{fig:numInteractions_Y10}
\end{figure}

Recall that we use \textit{leave-one-out-split} to partition the dataset into training and test sets. That is, we treat each user's last interaction as test instance while the remaining interactions that happened before the test instance are training instances. The number of user's interactions in the training set can be an indicator of user loyalty. For simplicity, we rank users by their number of interactions in training set. The users who rank among the top 50\% are \textit{loyal users}; the rest are \textit{non-loyal users}. The average number of interactions for both loyal users and non-loyal users can be found in Table~\ref{tab:distribution}. Note that, for SASRec and TiSASRec, only the most recent $n$ interactions of each user are used in training. That is because SASRec and TiSASRec are built on top of self-attention network which requires memory quadratic to sequence length of user interactions. Following the original papers of SASRec and TiSASRec, we tune $n$ as a hyperparameter. Hence, in Table~\ref{tab:distribution}, the average number of interactions for loyal users and non-loyal users appears to be different for SASRec, TiSASRec and the general recommenders - BPR, LightGCN and NeuMF. 

Figure~\ref{fig:numInteractions_Y10} plots the recommendation accuracy by HR@10 and NDCG@10 for loyal and non-loyal users that are defined by number of interactions. For HR@10, all the five models give better recommendations to non-loyal users on MovieLens-25M, Yelp, and Amazon-music. On Amazon-electronics, only  SASRec and TiSASRec offer better results for loyal users. As for NDCG@10, non-loyal users enjoy better recommendations than loyal users except for TiSASRec on Amazon-music as well as SASRec and TiSASRec on Amazon-electronic. Nevertheless, the overall trend suggests that \textit{non-loyal users enjoy better recommendations than loyal users}, particularly for general recommenders. This finding violates our intuition that more historical data leads to better learning of user's preference in recommendation. Hence, we hypothesize that not all historical interactions by a user are helpful for making ``recent'' recommendations. 

\subsection{Loyalty by Active Time Period}
\label{ssec:atp}

\begin{figure}
    \centering
\begin{subfigure}{\columnwidth}
        \centering
        \includegraphics[width=0.7\columnwidth]{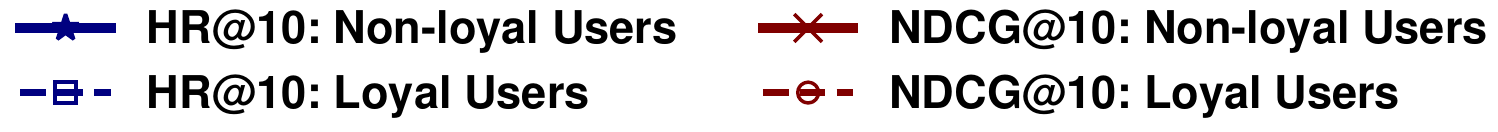}
        \Description{}
    \end{subfigure}
    \quad
    \begin{subfigure}{\columnwidth}
        \centering
        \includegraphics[width=\columnwidth]{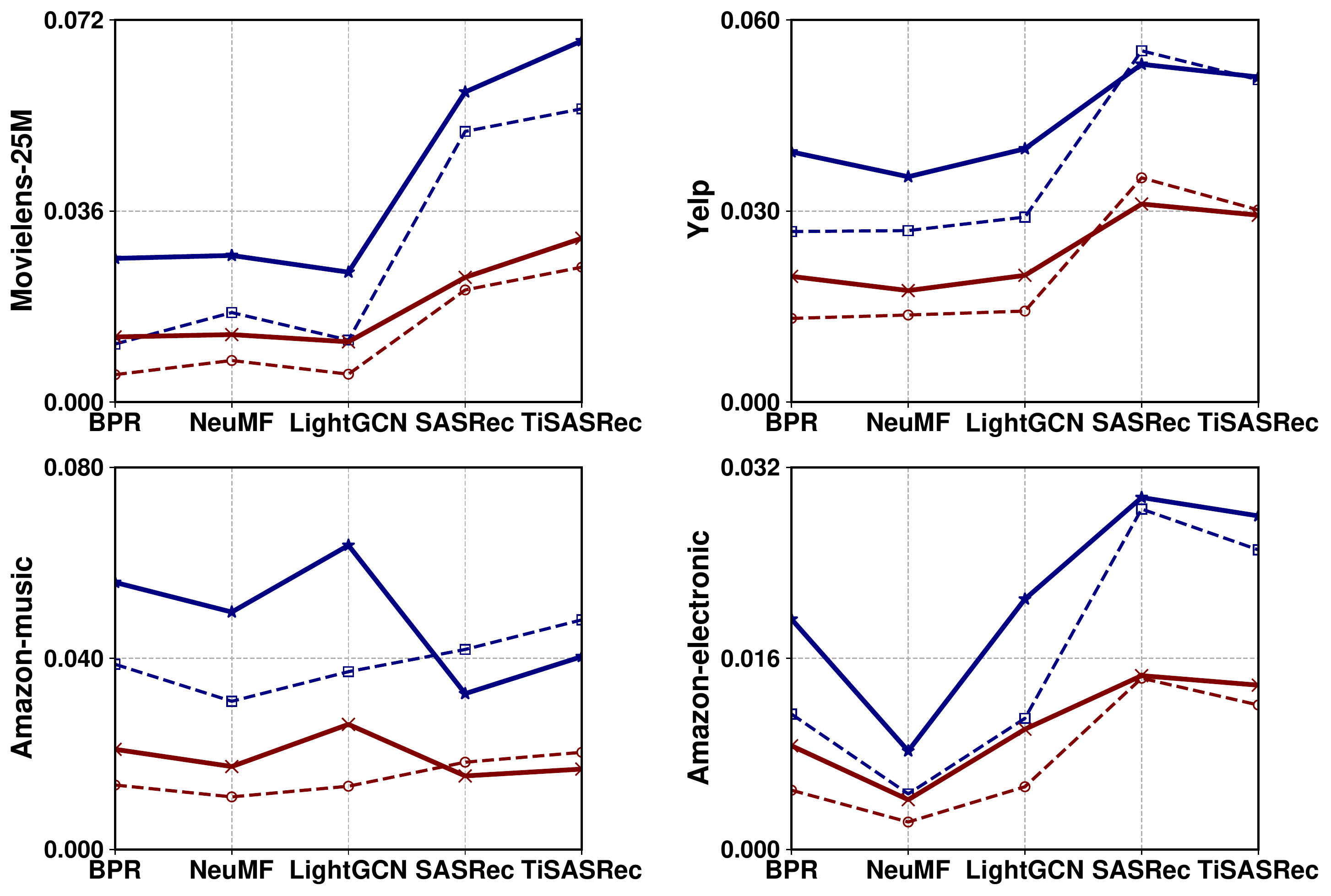}
    \end{subfigure}
        \caption{HR@10 and NDCG@10 of loyal and non-loyal users by \textbf{active time period}, in test year $Y10$}
    
    \label{fig:atp_Y10}
\end{figure}

Purely considering \textit{number of interactions} as a loyalty measure is not sufficient. A user can have many interactions within a short time period.
In light of this, we introduce a new loyalty indicator: Active Time Period (ATP), which is the number of days between a user's first interaction and last interaction. If a user has a long active time period, then the user has a long-term engagement with the system, and is a loyal user.

In our experiments, ATP is the number of days between a user's first interaction in the training set and the time of his/her test instance. Loyal users have ATP longer than the median ATP value.  Figure~\ref{fig:atp_Y10} plots HR@10 and NDCG@10 of loyal and non-loyal users indicated by active time period.

In general, non-loyal users, who have shorter active time period, enjoy better HR@10 and NDCG@10 than loyal users. A loyal user, by active time period, is a user who has started interacting with the system since a long time ago. On the one hand, the system has a better chance to capture the user's long-term interest. On the other hand, interactions that happened long time ago may not necessary represent the user's current preference. In Figure~\ref{fig:atp_Y10}, our results suggest mostly the latter on multiple datasets. We also observe that loyal and non-loyal users enjoy comparable recommendation accuracy by SASRec and TiSASRec on four datasets. That is because SASRec and TiSASRec are designed to treat recent interactions and old interactions differently for each individual user.

So far, our results show that preference of loyal users may not be well predicted, especially for general recommenders which treat all training instances equally. We hypothesize that this observation is attributed to the outdated interactions that happened a long time ago. These interactions cannot reflect a user's current interest. Hence, it motivates us to investigate another time factor: recency. 

\subsection{Recency}
\label{ssec:recency}

\begin{figure}
    \centering
    \begin{subfigure}{\columnwidth}
        \centering
        \includegraphics[width=0.7\columnwidth]{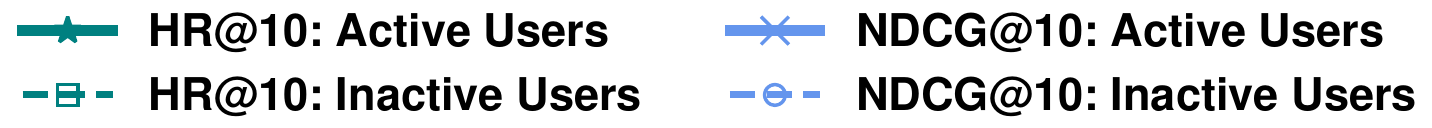}
        \Description{}
    \end{subfigure}
    \quad
    \begin{subfigure}{\columnwidth}
        \centering
        \includegraphics[width=\columnwidth]{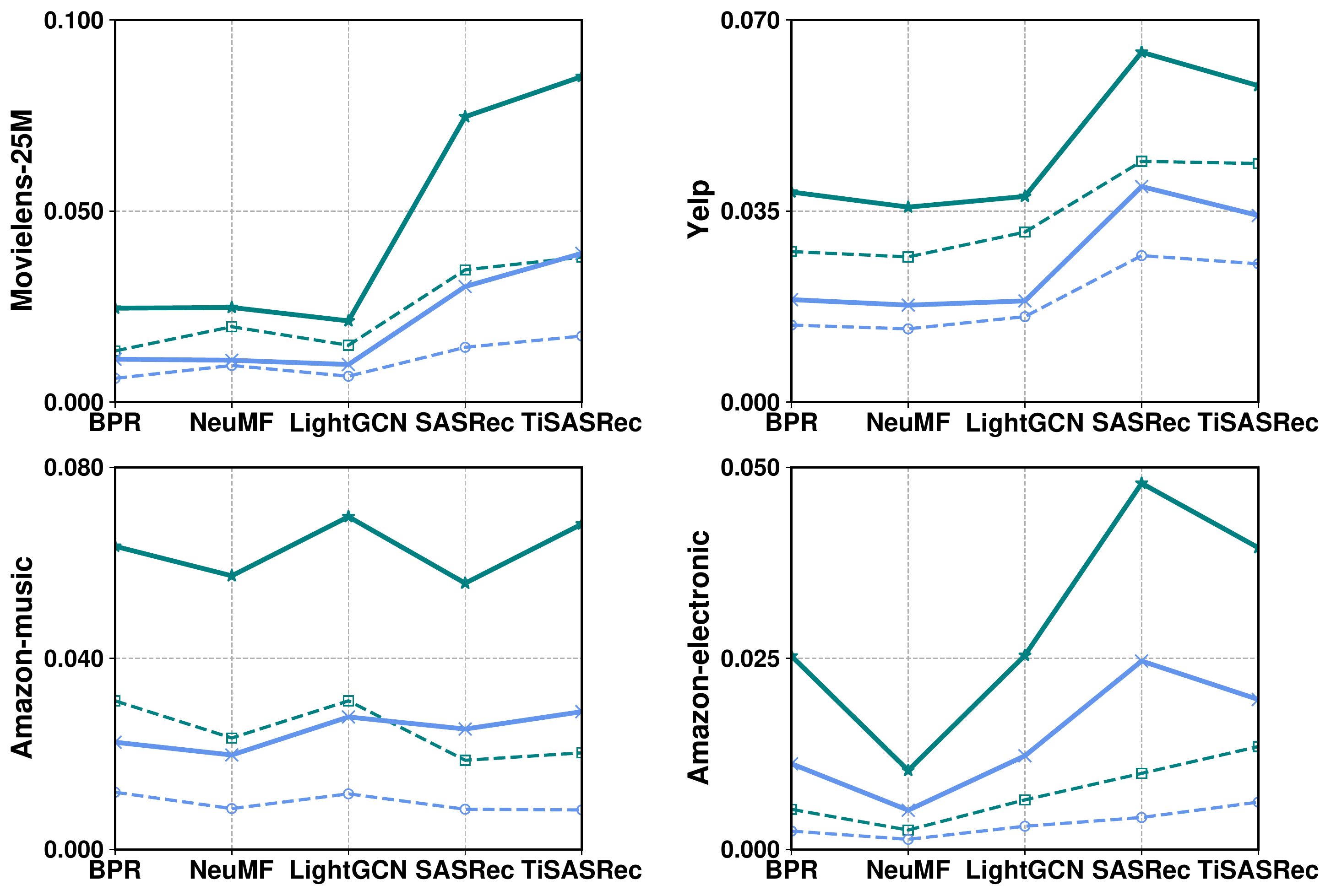}
    \end{subfigure}
    \caption{HR@10 and NDCG@10 for users grouped by recency of previous interaction before test instance.}
    \label{fig:recency_Y10}
\end{figure}

We define recency by the number of days between a user's test instance and his/her previous interaction just before the test instance. Again, we rank all users and take the $50^{th}$ percentile recency as threshold. We name the users with short recency as \textit{active users} and the remaining as \textit{inactive users}. As shown in Table~\ref{tab:distribution}, average recency for active users and inactive users are of the same values for all recommenders. Regardless of whether the models are general recommenders or sequential recommenders, recency value is not affected because it is calculated based on the time of the last interaction and the time of the test instance. 

Figure~\ref{fig:recency_Y10} plots HR@10 and NDCG@10 scores for both user groups. All recommenders, both general and sequential recommenders, deliver better recommendation results for active users, by both HR@10 and NDCG@10 measures. This finding provides strong evidence that more recent interactions matter more. This finding also partially explains the previous findings which suggest that loyal users suffer from poorer recommendations than non-loyal users. Moreover, we note that all recommenders, regardless of whether they consider time information in their model design, show similar trends on the recency factor. Even time-aware recommenders, SASRec and TiSASRec, do not consider the recency factor because ``time'' is considered only during model training, but not when making recommendations at the test time. In fact, SASRec and TiSASRec both consider local timeline specific to a user (\ie a user's interaction sequence) but not the global timeline. 
Based on the finding on recency, we argue that recommender shall consider recent interactions dynamically, with respect to the time point when making recommendations, along the global timeline.

\section{Conclusion}
\label{ssec:conclusion}

In this paper, through experiments, we show that \textbf{time matters} in recommendation. Here, we refer time with respect to the ``global timeline'' instead of ``local timeline'' where the latter is specific to a user. Having many interactions or having a long active time may adversely affect recommendation accuracy. Instead, recommenders give better results if users have recent interactions. On the other hand, recent interactions change continuously along the global timeline. Hence, we call for a revisit of model design in recommender system. The model should be able to capture the dynamics along the global timeline, thus to update user preference with the most recent context.

\bibliographystyle{ACM-Reference-Format}
\bibliography{reference}


\begin{thebibliography}{26}


\ifx \showCODEN    \undefined \def \showCODEN     #1{\unskip}     \fi
\ifx \showDOI      \undefined \def \showDOI       #1{#1}\fi
\ifx \showISBNx    \undefined \def \showISBNx     #1{\unskip}     \fi
\ifx \showISBNxiii \undefined \def \showISBNxiii  #1{\unskip}     \fi
\ifx \showISSN     \undefined \def \showISSN      #1{\unskip}     \fi
\ifx \showLCCN     \undefined \def \showLCCN      #1{\unskip}     \fi
\ifx \shownote     \undefined \def \shownote      #1{#1}          \fi
\ifx \showarticletitle \undefined \def \showarticletitle #1{#1}   \fi
\ifx \showURL      \undefined \def \showURL       {\relax}        \fi
\providecommand\bibfield[2]{#2}
\providecommand\bibinfo[2]{#2}
\providecommand\natexlab[1]{#1}
\providecommand\showeprint[2][]{arXiv:#2}

\bibitem[Bai et~al\mbox{.}(2019)]%
        {CTRec}
\bibfield{author}{\bibinfo{person}{Ting Bai}, \bibinfo{person}{Lixin Zou},
  \bibinfo{person}{Wayne~Xin Zhao}, \bibinfo{person}{Pan Du},
  \bibinfo{person}{Weidong Liu}, \bibinfo{person}{Jian{-}Yun Nie}, {and}
  \bibinfo{person}{Ji{-}Rong Wen}.} \bibinfo{year}{2019}\natexlab{}.
\newblock \showarticletitle{CTRec: {A} Long-Short Demands Evolution Model for
  Continuous-Time Recommendation}. In \bibinfo{booktitle}{\emph{SIGIR}}.
  \bibinfo{publisher}{{ACM}}, \bibinfo{pages}{675--684}.
\newblock


\bibitem[Beel(2017)]%
        {timeEval}
\bibfield{author}{\bibinfo{person}{J{\"{o}}ran Beel}.}
  \bibinfo{year}{2017}\natexlab{}.
\newblock \showarticletitle{It's Time to Consider "Time" when Evaluating
  Recommender-System Algorithms [Proposal]}.
\newblock \bibinfo{journal}{\emph{CoRR}}  \bibinfo{volume}{abs/1708.08447}
  (\bibinfo{year}{2017}).
\newblock


\bibitem[Chakraborty et~al\mbox{.}(2017)]%
        {recencyReleancy}
\bibfield{author}{\bibinfo{person}{Abhijnan Chakraborty},
  \bibinfo{person}{Saptarshi Ghosh}, \bibinfo{person}{Niloy Ganguly}, {and}
  \bibinfo{person}{Krishna~P. Gummadi}.} \bibinfo{year}{2017}\natexlab{}.
\newblock \showarticletitle{Optimizing the Recency-Relevancy Trade-off in
  Online News Recommendations}. In \bibinfo{booktitle}{\emph{{WWW}}}.
  \bibinfo{publisher}{{ACM}}, \bibinfo{pages}{837--846}.
\newblock


\bibitem[Faggioli et~al\mbox{.}(2020)]%
        {recencyCF}
\bibfield{author}{\bibinfo{person}{Guglielmo Faggioli}, \bibinfo{person}{Mirko
  Polato}, {and} \bibinfo{person}{Fabio Aiolli}.}
  \bibinfo{year}{2020}\natexlab{}.
\newblock \showarticletitle{Recency Aware Collaborative Filtering for Next
  Basket Recommendation}. In \bibinfo{booktitle}{\emph{{UMAP}}}.
  \bibinfo{publisher}{{ACM}}, \bibinfo{pages}{80--87}.
\newblock


\bibitem[Filipovic et~al\mbox{.}(2021)]%
        {temporalContext}
\bibfield{author}{\bibinfo{person}{Milena Filipovic}, \bibinfo{person}{Blagoj
  Mitrevski}, \bibinfo{person}{Diego Antognini}, \bibinfo{person}{Emma~Lejal
  Glaude}, \bibinfo{person}{Boi Faltings}, {and} \bibinfo{person}{Claudiu
  Musat}.} \bibinfo{year}{2021}\natexlab{}.
\newblock \showarticletitle{Modeling Online Behavior in Recommender Systems:
  The Importance of Temporal Context}. In \bibinfo{booktitle}{\emph{{RecSys}}}
  \emph{(\bibinfo{series}{{CEUR} Workshop Proceedings},
  Vol.~\bibinfo{volume}{2955})}. \bibinfo{publisher}{CEUR-WS.org}.
\newblock


\bibitem[Hansen et~al\mbox{.}(2020)]%
        {contextMusicRec}
\bibfield{author}{\bibinfo{person}{Casper Hansen}, \bibinfo{person}{Christian
  Hansen}, \bibinfo{person}{Lucas Maystre}, \bibinfo{person}{Rishabh Mehrotra},
  \bibinfo{person}{Brian Brost}, \bibinfo{person}{Federico Tomasi}, {and}
  \bibinfo{person}{Mounia Lalmas}.} \bibinfo{year}{2020}\natexlab{}.
\newblock \showarticletitle{Contextual and Sequential User Embeddings for
  Large-Scale Music Recommendation}. In \bibinfo{booktitle}{\emph{RecSys 2020:
  Fourteenth {ACM} Conference on Recommender Systems, Virtual Event, Brazil,
  September 22-26, 2020}}. \bibinfo{publisher}{{ACM}}, \bibinfo{pages}{53--62}.
\newblock


\bibitem[He et~al\mbox{.}(2020)]%
        {lightgcn}
\bibfield{author}{\bibinfo{person}{Xiangnan He}, \bibinfo{person}{Kuan Deng},
  \bibinfo{person}{Xiang Wang}, \bibinfo{person}{Yan Li},
  \bibinfo{person}{Yong{-}Dong Zhang}, {and} \bibinfo{person}{Meng Wang}.}
  \bibinfo{year}{2020}\natexlab{}.
\newblock \showarticletitle{LightGCN: Simplifying and Powering Graph
  Convolution Network for Recommendation}. In
  \bibinfo{booktitle}{\emph{{SIGIR}}}. \bibinfo{publisher}{{ACM}},
  \bibinfo{pages}{639--648}.
\newblock


\bibitem[He et~al\mbox{.}(2017)]%
        {neuMF}
\bibfield{author}{\bibinfo{person}{Xiangnan He}, \bibinfo{person}{Lizi Liao},
  \bibinfo{person}{Hanwang Zhang}, \bibinfo{person}{Liqiang Nie},
  \bibinfo{person}{Xia Hu}, {and} \bibinfo{person}{Tat{-}Seng Chua}.}
  \bibinfo{year}{2017}\natexlab{}.
\newblock \showarticletitle{Neural Collaborative Filtering}. In
  \bibinfo{booktitle}{\emph{{WWW}}}. \bibinfo{publisher}{{ACM}},
  \bibinfo{pages}{173--182}.
\newblock


\bibitem[Hidasi et~al\mbox{.}(2016)]%
        {GRU4REC}
\bibfield{author}{\bibinfo{person}{Bal{\'{a}}zs Hidasi},
  \bibinfo{person}{Alexandros Karatzoglou}, \bibinfo{person}{Linas Baltrunas},
  {and} \bibinfo{person}{Domonkos Tikk}.} \bibinfo{year}{2016}\natexlab{}.
\newblock \showarticletitle{Session-based Recommendations with Recurrent Neural
  Networks}. In \bibinfo{booktitle}{\emph{{ICLR}}}.
\newblock


\bibitem[Huang et~al\mbox{.}(2018)]%
        {knowledgeMemory}
\bibfield{author}{\bibinfo{person}{Jin Huang}, \bibinfo{person}{Wayne~Xin
  Zhao}, \bibinfo{person}{Hongjian Dou}, \bibinfo{person}{Ji{-}Rong Wen}, {and}
  \bibinfo{person}{Edward~Y. Chang}.} \bibinfo{year}{2018}\natexlab{}.
\newblock \showarticletitle{Improving Sequential Recommendation with
  Knowledge-Enhanced Memory Networks}. In \bibinfo{booktitle}{\emph{SIGIR}}.
  \bibinfo{publisher}{{ACM}}, \bibinfo{pages}{505--514}.
\newblock


\bibitem[Jeunen(2019)]%
        {revisitOffline}
\bibfield{author}{\bibinfo{person}{Olivier Jeunen}.}
  \bibinfo{year}{2019}\natexlab{}.
\newblock \showarticletitle{Revisiting offline evaluation for implicit-feedback
  recommender systems}. In \bibinfo{booktitle}{\emph{{RecSys}}}.
  \bibinfo{publisher}{{ACM}}, \bibinfo{pages}{596--600}.
\newblock


\bibitem[Ji et~al\mbox{.}(2022)]%
        {dataLeakage}
\bibfield{author}{\bibinfo{person}{Yitong Ji}, \bibinfo{person}{Aixin Sun},
  \bibinfo{person}{Jie Zhang}, {and} \bibinfo{person}{Chenliang Li}.}
  \bibinfo{year}{2022}\natexlab{}.
\newblock \showarticletitle{A Critical Study on Data Leakage in Recommender
  System Offline Evaluation}.
\newblock \bibinfo{journal}{\emph{CoRR}}  \bibinfo{volume}{abs/2010.11060}
  (\bibinfo{year}{2022}).
\newblock


\bibitem[Kang and McAuley(2018)]%
        {sasRec}
\bibfield{author}{\bibinfo{person}{Wang{-}Cheng Kang} {and}
  \bibinfo{person}{Julian~J. McAuley}.} \bibinfo{year}{2018}\natexlab{}.
\newblock \showarticletitle{Self-Attentive Sequential Recommendation}. In
  \bibinfo{booktitle}{\emph{{ICDM}}}. \bibinfo{publisher}{{IEEE} Computer
  Society}, \bibinfo{pages}{197--206}.
\newblock


\bibitem[Krichene and Rendle(2020)]%
        {sampledMetrics}
\bibfield{author}{\bibinfo{person}{Walid Krichene} {and}
  \bibinfo{person}{Steffen Rendle}.} \bibinfo{year}{2020}\natexlab{}.
\newblock \showarticletitle{On Sampled Metrics for Item Recommendation}. In
  \bibinfo{booktitle}{\emph{{KDD}}}. \bibinfo{publisher}{{ACM}},
  \bibinfo{pages}{1748--1757}.
\newblock


\bibitem[Larrain et~al\mbox{.}(2015)]%
        {goodTimeBadTime}
\bibfield{author}{\bibinfo{person}{Santiago Larrain},
  \bibinfo{person}{Christoph Trattner}, \bibinfo{person}{Denis Parra},
  \bibinfo{person}{Eduardo Graells{-}Garrido}, {and} \bibinfo{person}{Kjetil
  N{\o}rv{\aa}g}.} \bibinfo{year}{2015}\natexlab{}.
\newblock \showarticletitle{Good Times Bad Times: {A} Study on Recency Effects
  in Collaborative Filtering for Social Tagging}. In
  \bibinfo{booktitle}{\emph{{RecSys}}}. \bibinfo{publisher}{{ACM}},
  \bibinfo{pages}{269--272}.
\newblock


\bibitem[Li et~al\mbox{.}(2017)]%
        {NARM}
\bibfield{author}{\bibinfo{person}{Jing Li}, \bibinfo{person}{Pengjie Ren},
  \bibinfo{person}{Zhumin Chen}, \bibinfo{person}{Zhaochun Ren},
  \bibinfo{person}{Tao Lian}, {and} \bibinfo{person}{Jun Ma}.}
  \bibinfo{year}{2017}\natexlab{}.
\newblock \showarticletitle{Neural Attentive Session-based Recommendation}. In
  \bibinfo{booktitle}{\emph{{CIKM}}}. \bibinfo{publisher}{{ACM}},
  \bibinfo{pages}{1419--1428}.
\newblock


\bibitem[Li et~al\mbox{.}(2020)]%
        {timeSasRec}
\bibfield{author}{\bibinfo{person}{Jiacheng Li}, \bibinfo{person}{Yujie Wang},
  {and} \bibinfo{person}{Julian~J. McAuley}.} \bibinfo{year}{2020}\natexlab{}.
\newblock \showarticletitle{Time Interval Aware Self-Attention for Sequential
  Recommendation}. In \bibinfo{booktitle}{\emph{{WSDM}}}.
  \bibinfo{publisher}{{ACM}}, \bibinfo{pages}{322--330}.
\newblock


\bibitem[Liu et~al\mbox{.}(2018)]%
        {STAMP}
\bibfield{author}{\bibinfo{person}{Qiao Liu}, \bibinfo{person}{Yifu Zeng},
  \bibinfo{person}{Refuoe Mokhosi}, {and} \bibinfo{person}{Haibin Zhang}.}
  \bibinfo{year}{2018}\natexlab{}.
\newblock \showarticletitle{{STAMP:} Short-Term Attention/Memory Priority Model
  for Session-based Recommendation}. In \bibinfo{booktitle}{\emph{{KDD}}}.
  \bibinfo{publisher}{{ACM}}, \bibinfo{pages}{1831--1839}.
\newblock


\bibitem[Peska and Vojt{\'{a}}s(2020)]%
        {offlineOnline}
\bibfield{author}{\bibinfo{person}{Ladislav Peska} {and} \bibinfo{person}{Peter
  Vojt{\'{a}}s}.} \bibinfo{year}{2020}\natexlab{}.
\newblock \showarticletitle{Off-line vs. On-line Evaluation of Recommender
  Systems in Small E-commerce}. In \bibinfo{booktitle}{\emph{{HT}}}.
  \bibinfo{publisher}{{ACM}}, \bibinfo{pages}{291--300}.
\newblock


\bibitem[Rendle et~al\mbox{.}(2009)]%
        {BPR}
\bibfield{author}{\bibinfo{person}{Steffen Rendle}, \bibinfo{person}{Christoph
  Freudenthaler}, \bibinfo{person}{Zeno Gantner}, {and} \bibinfo{person}{Lars
  Schmidt{-}Thieme}.} \bibinfo{year}{2009}\natexlab{}.
\newblock \showarticletitle{{BPR:} Bayesian Personalized Ranking from Implicit
  Feedback}. In \bibinfo{booktitle}{\emph{{UAI}}}. \bibinfo{publisher}{{AUAI}
  Press}, \bibinfo{pages}{452--461}.
\newblock


\bibitem[Sun et~al\mbox{.}(2020)]%
        {evaluatingRigorously}
\bibfield{author}{\bibinfo{person}{Zhu Sun}, \bibinfo{person}{Di Yu},
  \bibinfo{person}{Hui Fang}, \bibinfo{person}{Jie Yang},
  \bibinfo{person}{Xinghua Qu}, \bibinfo{person}{Jie Zhang}, {and}
  \bibinfo{person}{Cong Geng}.} \bibinfo{year}{2020}\natexlab{}.
\newblock \showarticletitle{Are We Evaluating Rigorously? Benchmarking
  Recommendation for Reproducible Evaluation and Fair Comparison}. In
  \bibinfo{booktitle}{\emph{{RecSys}}}. \bibinfo{publisher}{{ACM}},
  \bibinfo{pages}{23--32}.
\newblock


\bibitem[Tanjim et~al\mbox{.}(2020)]%
        {intentRec}
\bibfield{author}{\bibinfo{person}{Md.~Mehrab Tanjim}, \bibinfo{person}{Congzhe
  Su}, \bibinfo{person}{Ethan Benjamin}, \bibinfo{person}{Diane Hu},
  \bibinfo{person}{Liangjie Hong}, {and} \bibinfo{person}{Julian~J. McAuley}.}
  \bibinfo{year}{2020}\natexlab{}.
\newblock \showarticletitle{Attentive Sequential Models of Latent Intent for
  Next Item Recommendation}. In \bibinfo{booktitle}{\emph{{WWW}}}.
  \bibinfo{publisher}{{ACM} / {IW3C2}}, \bibinfo{pages}{2528--2534}.
\newblock


\bibitem[Wang et~al\mbox{.}(2020)]%
        {seqHypergraph}
\bibfield{author}{\bibinfo{person}{Jianling Wang}, \bibinfo{person}{Kaize
  Ding}, \bibinfo{person}{Liangjie Hong}, \bibinfo{person}{Huan Liu}, {and}
  \bibinfo{person}{James Caverlee}.} \bibinfo{year}{2020}\natexlab{}.
\newblock \showarticletitle{Next-item Recommendation with Sequential
  Hypergraphs}. In \bibinfo{booktitle}{\emph{{SIGIR}}}.
  \bibinfo{publisher}{{ACM}}, \bibinfo{pages}{1101--1110}.
\newblock


\bibitem[Ye et~al\mbox{.}(2020)]%
        {timeMatters}
\bibfield{author}{\bibinfo{person}{Wenwen Ye}, \bibinfo{person}{Shuaiqiang
  Wang}, \bibinfo{person}{Xu Chen}, \bibinfo{person}{Xuepeng Wang},
  \bibinfo{person}{Zheng Qin}, {and} \bibinfo{person}{Dawei Yin}.}
  \bibinfo{year}{2020}\natexlab{}.
\newblock \showarticletitle{Time Matters: Sequential Recommendation with
  Complex Temporal Information}. In \bibinfo{booktitle}{\emph{{SIGIR}}}.
  \bibinfo{publisher}{{ACM}}, \bibinfo{pages}{1459--1468}.
\newblock


\bibitem[Zhao et~al\mbox{.}(2020)]%
        {Caper}
\bibfield{author}{\bibinfo{person}{Guoshuai Zhao}, \bibinfo{person}{Zhidan
  Liu}, \bibinfo{person}{Yulu Chao}, {and} \bibinfo{person}{Xueming Qian}.}
  \bibinfo{year}{2020}\natexlab{}.
\newblock \showarticletitle{CAPER: Context-Aware Personalized Emoji
  Recommendation}.
\newblock \bibinfo{journal}{\emph{{IEEE} Trans. Knowl. Data Eng.}}
  \bibinfo{volume}{32}.
\newblock


\bibitem[Zhao et~al\mbox{.}(2021)]%
        {Recbole}
\bibfield{author}{\bibinfo{person}{Wayne~Xin Zhao}, \bibinfo{person}{Shanlei
  Mu}, \bibinfo{person}{Yupeng Hou}, \bibinfo{person}{Zihan Lin},
  \bibinfo{person}{Yushuo Chen}, \bibinfo{person}{Xingyu Pan},
  \bibinfo{person}{Kaiyuan Li}, \bibinfo{person}{Yujie Lu},
  \bibinfo{person}{Hui Wang}, \bibinfo{person}{Changxin Tian},
  \bibinfo{person}{Yingqian Min}, \bibinfo{person}{Zhichao Feng},
  \bibinfo{person}{Xinyan Fan}, \bibinfo{person}{Xu Chen},
  \bibinfo{person}{Pengfei Wang}, \bibinfo{person}{Wendi Ji},
  \bibinfo{person}{Yaliang Li}, \bibinfo{person}{Xiaoling Wang}, {and}
  \bibinfo{person}{Ji-Rong Wen}.} \bibinfo{year}{2021}\natexlab{}.
\newblock \showarticletitle{RecBole: Towards a Unified, Comprehensive and
  Efficient Framework for Recommendation Algorithms}. In
  \bibinfo{booktitle}{\emph{CIKM}}. \bibinfo{publisher}{{ACM}},
  \bibinfo{pages}{4653–4664}.
\newblock


\end{thebibliography}

\end{document}